\documentstyle[eqsecnum,aps]{revtex}

\newcommand{\w}{\wedge}
\newcommand{\be}{\begin{equation}}
\newcommand{\ee}{\end{equation}}
\newcommand{\bea}{\begin{eqnarray}}
\newcommand{\eea}{\end{eqnarray}}
\newcommand{\om}{\omega}
\def\sutep{\rlap{\lower2ex\hbox{$\,\tilde{}$}}\epsilon{}}
\begin{document}
\draft
\title{A note on 4-dimensional traversable wormholes\\
and energy conditions in higher dimensions}
\author{W. F. Kao\cite{byline1}}
\address{Institute of Physics,\\
National Chiao Tung University\\
Hsinchu, Taiwan. }
\author{Chopin Soo\cite{byline2}}
\address{Department of Physics,\\
National Cheng Kung University\\
Tainan 70101, Taiwan. }

\maketitle
\begin{abstract}
We show explicitly that traversable wormholes requiring exotic
matter in 4-dimensions nevertheless have acceptable stress-tensors
obeying reasonable energy conditions in higher dimensions if the
wormholes are regarded as being embedded in higher dimensional
space-times satisfying Einstein's field equations. From the
4-dimensional perspective, the existence of higher dimensions may
thus facilitate wormhole and time-machine constructions through
access to {\it exotic matter}.

\end{abstract}

\widetext

\section{Introduction}

   It has been argued that
  if traversable wormholes exists, time-machines can be
  constructed\cite{Yurtsever,Visser}. On the other hand, although quantum effects are
  capable of violating null, weak, strong and dominant energy conditions\footnote{see, for instance
  Chapter 12 of Ref.\cite{Visser} for a useful discussion on the various energy conditions.
  It is worth pointing out that these energy conditions are assumed in the derivations of some
  singularity theorems and laws of black hole thermodynamics.};
  it is also known that wormholes require exotic matter\cite{Thorne},
  and must lead to violations of reasonable, classical physical restrictions such as the
  null energy condition\cite{Hochberg,Visser1,Visser2,Visser3}.
  So it is not entirely clear that stable traversable wormholes can also be realized physically, and
  are not merely theoretical constructs. Recently, there has also been great
  interest in the existence of higher dimensional space-times. Since a lower
  dimensional manifold can readily be embedded in a higher
  dimension, if higher dimensions exist, what appears as a wormhole to a 4-dimensional
  observer may indeed correspond to a restricted path in a higher dimensional space-time
  with a classically acceptable higher-dimensional stress-tensor.
  In other words, the ``4-dimensional wormhole traveller" encounters
  no violation of reasonable energy conditions from the perspective of the higher
  dimensional space-time. In this short note we demonstrate explicitly that
  this is possible by choosing just a particular class of traversable 4-dimensional wormhole
  solutions\cite{Thorne}; and showing that it leads to reasonable stress-tensors
  which satisfy the strong and dominant energy conditions if the
  5-dimensional space-time in which the wormhole is embedded is
  regarded as a solution of the higher dimensional Einstein field equations.
  This raises the implication that if higher dimensions exist, even without resorting to
  instances of exotic stress-tensors from quantum effects, time machines and causality violations
  in 4-dimensions may be encountered via trajectories in higher dimensions which appear as wormhole traversals
  to 4-dimensional observers.

\section{Wormhole solutions, embedding and higher dimensional curvature
tensor}

We start by choosing a particular class of suitable traversable wormhole solutions. Our aim is not to
exhaustively examine all possible solutions, but to explicitly demonstrate that the higher dimensional
stress-tensor is physically reasonable. To wit, we consider the wormhole solution of Ref. \cite{Thorne} in
4-dimensions expressed as \be ds^2 =-e^{2\phi(r)}dt^2 + dr^2/[1-b(r)/r] + r^2(d\theta^2 + \sin^2\theta d\phi^2).
\ee It can be embedded in 5 dimensions by introducing the fifth coordinate $z$ such that \be dz/dr =
\pm[r/b(r)-1]^{-\frac{1}{2}}. \ee Thus if we regard the 5-dimensional metric as \be ds^2 = -e^{2\phi(r)}dt^2 +
dz^2 + dr^2 + r^2(d\theta^2 + \sin^2\theta d\phi^2), \ee a restricted part of the higher dimension subject to
Eq.(2.2) corresponds to the 4-dimensional wormhole metric of Eq.(2.1). The 5-dimensional vielbien 1-forms, $e^A
\,(A= 0,1,2,3,4),$ are consequently \be \{e^0 =e^\phi dt ,\, e^1= dz ,\, e^2= dr ,\, e^3 = r d\theta ,\, e^4=
r\sin\theta d\phi\}; \ee and the spin connection can be straightforwardly evaluated via the torsionless condition
\be de^A + \om^A\,_B \w e^B =0.\ee Our convention is $\eta_{AB} = {\rm diag.}(-1,+1,+1,+1,+1)$ for the
5-dimensional Minskowski metric. If we denote $\phi'(r) \equiv d\phi(r)/dr$, then the only nonvanishing Lorentz
components of the spin connection 1-form are \bea \om^0\,_2 &=&\phi' e^\phi dt \qquad \om^2\,_3 = -d\theta,\cr
\om^2\,_4 &=& -\sin\theta d\phi, \qquad \om^3\,_4 = -\cos\theta d\phi; \eea and the only nontrivial components of
the curvature 2-form, ${\cal R}^A\,_B = d\om^A\,_B + \om^A\,_C \w \om^C\,_B$, are \be {\cal R}^0\,_2 =[\phi"
+\phi'^2]e^\phi dr\w dt, \quad {\cal R}^0\,_3 = -\phi'e^\phi dt\w d\theta, \quad {\cal R}^0\,_4 =-\phi' e^\phi dt
\w \sin\theta d\phi. \ee We may expand the the curvature 2-form using ${\cal R}_{AB} \equiv
\frac{1}{2}R_{ABCD}e^C\w e^D$, and it can be readily shown that the only nonvanishing components of the
Riemann-Christoffel curvature, $R_{ABCD}$, are \be R^0\,_{202} =-(\phi" + \phi'^2), \qquad R^0\,_{303} =
R^0\,_{404} = -\phi'/r \,. \ee This yields the only nontrivial Ricci curvature components, $R_{AB} =
R^C\,_{ACB}$, as \be R_{22} = -(\phi" + \phi'^2), \quad R_{33}=R_{44}= -\phi'/r;\ee and the Ricci scalar
curvature, $R= R^A\,_A$, as \be R= -2(\phi" +\phi'^2) -4\phi'/r .\ee With these we can compute the Einstein
tensor \bea G_{AB} &=& R_{AB} -\frac{1}{2}\eta_{AB}R \cr &=& R_{AB} +\eta_{AB}[2\phi'/r + (\phi" + \phi'^2)].\eea
Gathering the previous results, the Einstein tensor is therefore diagonal, and has components \bea G_{00} &=& 0,
\quad G_{11} = 2\phi'/r +( \phi" + \phi'^2), \quad G_{22} = 2\phi'/r, \cr G_{33} &=& G_{44} = \phi'/r +(\phi"
+\phi'^2). \eea Assuming that Einstein's field equations for the higher dimension are satisfied i.e. \be G_{AB} =
\kappa^2_0 T_{AB},\ee where $\kappa_0$ is the 5-dimensional coupling; it follows that the 5-dimensional
energy-momentum or stress tensor must be \be T_{AB} = \frac{1}{\kappa^2_0}\left(\matrix{0&0&0&0&0\cr
0&2\phi'/r+(\phi"+ \phi'^2)&0&0&0\cr 0&0&2\phi'/r & 0&0\cr 0&0&0&\phi'/r+(\phi"+ \phi'^2) &0\cr 0&0&0&0&\phi'/r
+(\phi" + \phi'^2)}\right).\ee

\section{Energy conditions}

In order for the 4-dimensional metric of Eq.(2.1) to describe suitable traversable 4-dimensional wormhole
solutions, the parameters are known to be\cite{Thorne} \be \phi' =\frac{ -\kappa^2\tau(r) r^3 +
b(r)}{2r(r-b(r))},\ee where $\kappa^2 =8\pi G$ and $\tau$ is the tension per unit area in the r-direction of the
wormhole i.e. the negative of the radial pressure. For comparison, if we assume Einstein's theory, the
4-dimensional stress-tensor of the metric of Eq.(2.1) is\cite{Thorne} \be T_{ab} =
\frac{1}{\kappa^2}\left(\matrix{\rho=\frac{b'}{\kappa^2 r^2}&0&0&0\cr 0&-\tau&0&0\cr 0&0& p=\frac{r}{2}[(\rho
-\tau)\phi'-\tau']-\tau &0\cr 0&0&0& p}\right),\ee where $a,b =0,1,2,3$ denote the components in the unit basis
of the vierbien of Eq.(2.1); and $b'(r) \equiv db(r)/dr$. As explained in Ref.\cite{Thorne}, the flaring out
condition $(d^2r/dz^2)|_{r=b_0} =[(b(r) -b'(r)r)/(2b^2(r))]|_{b_0}
>0$ and $[(r-b(r))\phi'(r)]|_{b_0} =0$ at the throat ($r=b_0$) of
the wormhole implies $\tau(b_0) > \rho(b_0)$. Thus from the 4-dimensional perspective ``exotic matter" will be
required and the null energy condition,\footnote{see Chapter 12 of Ref.\cite{Visser} for a discussion of the
energy conditions and their equivalence to the relations used in this article.} \be \rho + p_i \geq 0 \quad
\forall \,i, \ee is obviously violated, at least at the throat of the wormhole. This implies a violation of the
weak energy condition which includes the statement $\rho \geq 0$ besides the null energy condition of (3.3). On
the other hand it is is easy to check there exist suitable choices of the wormhole shape function $b(r)$ and the
factor $\phi(r)$ controlling time dilation which do {\it not} violate suitable energy conditions for the
5-dimensional stress-tensor of Eq.(2.14). For instance, we may adopt $\phi(r)=0$, $b=$ positive constant and
$\tau(r) = b_0/(\kappa^2 r^3)$ - indeed these parameters were suggested in Ref. \cite{Thorne} to minimize exotic
material. This gives $\tau'(b_0)= -3/(\kappa^2 b^2_0).$ Consequently, for the 4-dimensional stress-tensor of
Eq.(3.2), the null energy condition and also the stronger statement of dominant energy condition, \be \rho \geq 0
\quad {\rm and} \quad \forall \,i \quad p_i \in [-\rho,\rho], \ee are violated; and for constant tension
$(\tau'=0)$ the strong energy condition, \be \forall \,i \quad \rho + p_i \geq 0 \quad {\rm and} \quad \rho +
\sum_i p_i \geq 0, \ee is also not satisfied. But from the perspective of the higher-dimensional theory, there is
no ``exotic matter" whatsoever as the strong and dominant energy conditions clearly hold for the 5-dimensional
stress-tensor of Eq.(2.14). In this particular instance, the higher dimensional embedding manifold is just the
Minkowski space-time obeying vacuum ($T_{AB} =0$ ) Einstein field equations in 5-dimensions.

  The results can be discussed in a more general setting.
  If our 4-dimensional manifold $M$ is described by a 3-brane with metric
  $q_{\mu\nu} = g_{\mu\nu} -n_\mu n_\nu$ in a 5-dimensional space-time with metric $g_{\mu\nu}$ and
  the normal to $M$ is $n_\mu$, then the metric $g_{\mu\nu}$ of a theory satisfying the 5-dimensional Einstein
  field equation of Eq.(2.13) will produce a 4-dimensional Einstein tensor of the form\cite{Shiromizu}
  \be G_{\mu\nu} = \frac{2\kappa^2_0}{3}
\left(T_{\rho\sigma}q^\rho_\mu q^\sigma_\nu +(T_{\rho\sigma}n^\rho n^\sigma-\frac{1}{4}T)q_{\mu\nu}\right) + K
K_{\mu\nu} -K^\sigma_\mu K_{\nu\sigma} -\frac{1}{2}q_{\mu\nu}(K^2-K^{\alpha\beta}K_{\alpha\beta}) -
C^\alpha\,_{\beta\rho\sigma}n_\alpha n^\rho q^\beta_\mu q^\sigma_\nu \ee where $T$ and $K$ are respectively the
trace of the higher dimensional stress-tensor and the extrinsic curvature $K_{\mu\nu}$, and
$C^\alpha\,_{\beta\rho\sigma}$ is the higher-dimensional Weyl or conformal tensor. We showed {\it explicitly}
that even when $T_{\mu\nu}$ obeys reasonable energy conditions, $G_{\mu\nu}$ can nevertheless be ``exotic".
Moreover, our $(M, q_{\mu\nu})$ corresponds specifically to an interesting class of traversable 4-dimensional
wormholes. From the 4-dimensional perspective, the existence of higher dimensions may thus facilitate wormhole
and time machine constructions through access to ``exotic matter" but may also entail perplexing chronology
violations via wormhole traversals.

\acknowledgments  The research for this work has been supported in
part by funds from the National Science Council of Taiwan under
grants nos. NSC91-2112-M-009-034, NSC90-2112-M-006-012 and
NSC91-2112-M-006-018.


\end{document}